\RequirePackage{fix-cm}
\documentclass[twocolumn,epjc3]{svjour3}  
\usepackage{graphicx} 
\usepackage{dcolumn}
\usepackage{bm}
\usepackage{subfigure}
\usepackage{float}
\usepackage{multirow}
\usepackage{booktabs}
\usepackage{amsmath}
\usepackage{cases}
\usepackage{latexsym,amssymb}
\usepackage[mathscr]{eucal}
\usepackage[switch]{lineno}
\usepackage{array}
\usepackage{booktabs}
\usepackage[bookmarks=true,
   colorlinks=true,
   linkcolor=blue,
   urlcolor=blue,
   citecolor=blue,
   bookmarks=true,
   hyperindex=true
]{hyperref}
\smartqed  
\usepackage{xcolor}

\smartqed  

\journalname{Eur. Phys. J. C}

\begin{document}

\title{Constraining dark-matter-admixed hyperon stars with multimessenger observations: Implications for $f$-mode oscillations
}


\author{Muqi Zhang\thanksref{addr1}
        \and
        Xia Zhou\thanksref{e1,addr1}
        \and
        Xueling Mu\thanksref{addr2}
        \and
        Zhongwen Feng\thanksref{addr1}
}

\thankstext{e1}{e-mail: zhoux@my.swjtu.edu.cn(corresponding author)} 


\institute{School of Physics and Astronomy,
China West Normal University, Nanchong 637009, China \label{addr1}
           \and
           College of Electronic Information and Electrical Engineering, Chengdu University, Chengdu, 610106, China \label{addr2}
}

\date{Received: date / Accepted: date}
\maketitle

\begin{abstract}
We investigate the stellar structure, tidal deformability, and nonradial $f$-mode oscillation frequencies of dark-matter-admixed hyperon stars, using multimessenger observations to constrain the model parameters. Within the relativistic mean-field framework, we consider six different hyperon-meson coupling schemes and adopt a single-fluid model incorporating fermionic dark matter. Our results show that variations in the hyperon-meson couplings primarily affect the equation of state at high densities and the maximum stellar mass. In the absence of dark matter, none of the six models can satisfy the mass-radius constraints of HESS J1731$-$347. The inclusion of dark matter allows all six models to satisfy these constraints within certain parameter ranges. The NICER mass-radius measurements exclude two of the six coupling schemes, while the maximum-mass requirement $M_{\max}\geq 2M_\odot$ further restricts the allowed range of dark matter Fermi momentum. The remaining four models also satisfy the tidal deformability constraints from GW170817. For the observationally allowed models, the quadrupolar $f$-mode frequencies generally increase with stellar mass and dark matter Fermi momentum, with the effects of dark matter becoming more pronounced in massive neutron stars. Differences among the hyperon-meson coupling schemes are small for canonical-mass neutron stars but reach the order $100~\mathrm{Hz}$ at higher masses. These findings demonstrate that multimessenger observations can effectively constrain the parameter space of dark-matter-admixed hyperon star models, while $f$-mode oscillations provide complementary dynamical information that may help distinguish between microscopic models with similar static properties.

\keywords{Neutron stars\and Hyperons\and Dark matter \and Multimessenger observations\and $f$-mode oscillations}
\end{abstract}

\section{Introduction}
\label{intro}
Neutron stars provide a unique astrophysical environment for probing strongly interacting matter at suprasaturation densities, while our current understanding of their interior composition and properties relies primarily on observations of their macroscopic characteristics. Electromagnetic and gravitational-wave observations provide constraints on neutron star masses, radii, and tidal deformabilities~\cite{Ozel:2016oaf,Abbott:2018exr,Miller:2026NICER}, while future asteroseismic and gravitational-wave observations may further reveal their oscillation properties~\cite{Andersson:1997rn}. These macroscopic properties are directly influenced by the composition of stellar matter and its equation of state (EOS), making them important observational probes of the microscopic physics of dense matter~\cite{Lattimer:2000nx,Oertel:2017EOS,Baym:2018HadronQuark,Tsang:2024NuclearNS}. However, existing observations do not uniquely determine the microscopic physics of neutron star interiors. Different microscopic degrees of freedom can modify the EOS in distinct ways while producing similar stellar structures and responses. Distinguishing between different microscopic mechanisms that lead to similar macroscopic behavior by exploiting observables with different sensitivities to stellar interiors therefore remains an important challenge in the study of dense matter~\cite{Burgio:2021vgk,Biswas:2025EOSMass}.

Dense matter may contain various non-nucleonic degrees of freedom, and hyperons are among the important candidates. As the baryon density increases, strange baryons may emerge in neutron star interiors, significantly affecting the EOS at high densities by modifying the particle composition and chemical equilibrium~\cite{Tolos:2020aln}. The appearance of hyperons is generally known to soften the EOS, thereby reducing the maximum mass that neutron stars can support and giving rise to the long-standing hyperon puzzle ~\cite{Chatterjee:2015pua,Zhu:2026HyperonPuzzle}. However, the degree of this softening is not fixed but depends strongly on hyperon interactions at high densities. Within the relativistic mean-field framework, these interactions are described through hyperon-meson couplings, whose coupling constants remain subject to substantial theoretical uncertainties\cite{Schaffner:1996Hyperon,Ding:2025Hyperon,Athira:2026SU3SU6}. Consequently, even with the same underlying nucleonic model, uncertainties in hyperon interactions can lead to a range of different predictions for stellar structure~\cite{Huang:2025Hyperon}. This raises the question of whether the variations in the EOS and stellar structure arising from different hyperon interactions are sufficient to simultaneously satisfy neutron star observations across different mass ranges~\cite{Kochankovski:2025HyperonMerger}.

The low-mass compact object HESS J1731$-$347 and massive pulsars provide complementary constraints for testing the consistency of hyperonic models across different mass ranges~\cite{Bauswein:2026Hyperons}. Mass and radius estimates based on X-ray spectral analysis and distance measurements suggest that HESS J1731$-$347, if interpreted as a neutron star, may correspond to a low-mass and relatively compact stellar configuration ~\cite{Doroshenko:2022HESS,Char:2026HESS,Huang:2024TensorHESS,Mariani:2024SlowHybridHESS,Tewari:2025HESS}. In contrast, massive pulsars such as PSR J0740+6620 require the EOS to provide sufficient pressure support at high densities to sustain neutron stars with masses of approximately $2M_\odot$~\cite{Fonseca:2021J0740,Miller:2021J0740,Riley:2021J0740}. Since hyperons generally have a significant impact on the matter composition and EOS only above a certain density threshold, their effects are typically more pronounced in massive neutron stars and relatively limited in low-mass stellar configurations~\cite{Bauswein:2026Hyperons}. This raises the question of whether varying hyperon interactions alone can simultaneously satisfy these two sets of observational constraints.

In addition to hyperons, dark matter may also serve as an additional microscopic degree of freedom that affects the structure of neutron stars. Higgs-coupled fermionic dark matter can modify the EOS , thereby altering the mass-radius relations and tidal properties of neutron stars ~\cite{Panotopoulos:2017Higgs,Das:2019Higgs,Iqbal:2026DM}. The effects of dark matter on stellar structure also depend on its particle mass, abundance, and interaction strength ~\cite{Barbat:2024CompactDM,Kumar:2026VectorPortal}. To investigate these effects, previous studies have employed single-fluid and two-fluid models to examine the structure, tidal deformability, and nonradial oscillations of dark-matter-admixed neutron stars and hyperon stars~\cite{Hong:2024MixedDM,Mu:2025HyperonDM,Mandal:2026DarkQuarkHESS}. Notably, although hyperon interactions and dark matter components have distinct microscopic origins, different hyperon coupling schemes and dark matter abundances may still yield similar mass-radius relations and other observable signatures, leading to observational degeneracies among different microscopic models~\cite{Shirke:2024DM,Shahrbaf:2026EOS,Biesdorf:2025Masquerading}. Consequently, a single observable is often insufficient to distinguish between microscopic models with similar macroscopic properties, making it necessary to combine multimessenger observations with different sensitivities to the stellar interior to further constrain the allowed parameter space~\cite{Mariani:2024Multimessenger,Kumar:2024DMConstraints,Cuceu:2025EOS}.

Among current observations, mass–radius measurements provide the most direct constraints on the static structure of neutron stars. As stellar mass increases, the central density generally also rises. Therefore, observations of neutron stars with different masses allow the EOS to be tested over a wider range of densities. For example, HESS J1731−347, PSR J0030+0451, and PSR J0740+6620 provide representative constraints on stellar structure in the low-mass, canonical-mass, and high-mass regimes, respectively~\cite{Doroshenko:2022HESS,Miller:2021J0740,Riley:2021J0740,Miller:2019J0030,Vinciguerra:2024J0030}. However, similar mass–radius relations may still correspond to different internal matter distributions, while the tidal deformability constraints from GW170817 provide additional information on the stellar interior~\cite{Abbott:2018exr,TheLIGOScientific:2017qsa,Ecker:2025Ringdown}. As a fundamental fluid oscillation mode of neutron stars, the $f$-mode is sensitive to the mean density, compactness, and internal matter composition~\cite{Benhar:2004Asteroseismology,Pradhan:2020Fmode,Pradhan:2022Fmode,Maiti:2026FmodeHyperons}, and its frequency can also be affected by the presence of dark matter~\cite{Das:2021Fmode,Kumar:2025Asteroseismology,Sen:2024DarkFmode,Flores:2024DMAsteroseismology,Thakur:2025DMSigmaCut}. Therefore, predicting the  $f$-mode oscillation frequencies of models consistent with existing observational constraints can provide dynamical information complementary to mass–radius relations and tidal deformability, offering a theoretical basis for probing the internal composition of neutron stars through future asteroseismic and gravitational-wave observations~\cite{Gittins:2026TidalResonances}.

Motivated by these considerations, we consider different hyperon-meson coupling schemes based on a common nucleonic model and incorporate fermionic dark matter coupled to ordinary matter through the Higgs field within a single-fluid framework. The effects of uncertainties in hyperon interactions on neutron star structure are first investigated. The allowed dark matter parameter space for each model is then progressively determined by combining observational constraints from HESS J1731$-$347, NICER, massive pulsars, and GW170817. On this basis, the nonradial $f$-mode oscillation frequencies of the models consistent with existing observations are calculated to compare their dynamical properties and assess their potential observational significance.

The paper is organized as follows. Section~\ref{sec2} presents the theoretical model of neutron stars containing hyperons and fermionic DM, including the baryonic matter, hyperon-meson couplings, the DM component, and the stellar structure equations. Section~\ref{sec3} first examines the effects of different hyperon interactions and DM components on neutron-star properties, then constrains the model parameter space using mass-radius and tidal-deformability observations, and further analyzes the nonradial $f$-mode oscillation frequencies of the remaining models. Section~\ref{sec4} summarizes the main results.

\section{Theoretical model}
\label{sec2}
\subsection{Baryonic matter model}
\label{sec2-1}
To investigate the effects of hyperonic degrees of freedom on the composition and EOS of high-density matter in neutron stars, we employ a relativistic mean-field (RMF) model including hyperons to describe baryonic matter, considering the full baryon octet $B=\{n,p,\Lambda,\Sigma^-,\Sigma^0,\Sigma^+,\Xi^-,\Xi^0\}$. Within the RMF framework, interactions among baryons are mediated by meson mean fields, including the $\sigma$, $\omega$, $\rho$, $\sigma^*$, and $\phi$ mesons. The $\sigma$ and $\omega$ mesons describe scalar attraction and vector repulsion, respectively, while the $\rho$ meson, as a vector-isovector meson, accounts for the isospin dependence of baryonic interactions. The strange mesons $\sigma^*$ and $\phi$ are primarily introduced to describe hyperon-hyperon interactions. The corresponding baryonic and mesonic Lagrangian densities are given by~\cite{Schaffner:1996Hyperon,Walecka:1974,Serot:1992}
\begin{align}
\label{eq51}
\mathcal{L}_{B} &=\sum_{B}\bar{\psi}_{B}
\left[i\gamma_{\mu}\partial^{\mu}
-M_B+g_{\sigma B}\sigma +g_{\sigma^*B}\sigma^*\right.
\nonumber \\
&\left.-g_{\omega  B}\gamma_{\mu}\omega^{\mu}-g_{\rho B}\gamma_{\mu}{I}_{B}\cdot{\rho}^{\mu} -g_{\phi B}\gamma_{\mu}\phi^{\mu}\right]\psi_B,
\end{align}
\begin{align}
\mathcal{L}_{M}&=\frac{1}{2}\partial_{\mu}\sigma\partial^{\mu}\sigma-\frac{1}{2}m_{\sigma}^{2}\sigma^{2}-\frac{1}{3}g_{2}\sigma^{3}-\frac{1}{4}g_{3}\sigma^{4}
\nonumber\\
&-\frac{1} {4}\omega_{\mu\nu}\omega^{\mu\nu}+\frac{1}{2}m_{\omega}^{2}\omega_{\mu}\omega^{\mu}-\frac{1}{4}\boldsymbol{\rho}_{\mu\nu}\!\cdot\!\boldsymbol{\rho}^{\mu\nu}
\nonumber\\
&+\frac{1}{2}m_{\rho}^{2}\boldsymbol{\rho}_{\mu}\!\cdot\!\boldsymbol{\rho}^{\mu}+\Lambda_{v}g_{\omega}^{2}g_{\rho}^{2}(\omega_{\mu}\omega^{\mu})(\boldsymbol{\rho}_{\nu}\!\cdot\!\boldsymbol{\rho}^{\nu})
\nonumber\\
&+\frac{1}{2}\partial_{\mu}\sigma^{*}\partial^{\mu}\sigma^{*}-\frac{1}{2}m_{\sigma^{*}}^{2}\sigma^{*2}-\frac{1}{4}\phi_{\mu\nu}\phi^{\mu\nu}
\nonumber \\
&+\frac{1}{2}m_{\phi}^{2}\phi_{\mu}\phi^{\mu}.
\end{align}
Here, $M_B$ and $\psi_B$ denote the vacuum mass and Dirac field of baryon $B$, respectively. The constants $g_{\sigma B}$, $g_{\omega B}$, $g_{\rho B}$, $g_{\sigma^* B}$, and $g_{\phi B}$ are the corresponding baryon-meson coupling constants, and $\boldsymbol{I}_B$ is the baryon isospin operator. The meson masses are denoted by $m_i$ $(i=\sigma,\omega,\rho,\sigma^*,\phi)$. The parameters $g_2$ and $g_3$ describe the nonlinear self-interactions of the $\sigma$ field, and $\Lambda_v$ is the $\omega$-$\rho$ mixing coupling constant. We use $g_\omega\equiv g_{\omega N}$ and $g_\rho\equiv g_{\rho N}$ for the couplings of nucleons to the $\omega$ and $\rho$ mesons, respectively. The field-strength tensors are defined by
\begin{subequations}
\begin{align}
\omega_{\mu\nu}&=\partial_\mu\omega_\nu- \partial_\nu\omega_\mu, 
\\
\phi_{\mu\nu}&=\partial_\mu\phi_\nu-\partial_\nu\phi_\mu, 
\\
\boldsymbol{\rho}_{\mu\nu}&=\partial_\mu\boldsymbol{\rho}_\nu-\partial_\nu\boldsymbol{\rho}_\mu.
\end{align}
\end{subequations}
To describe neutron star matter under the conditions of $\beta$ equilibrium and charge neutrality, electrons and muons must also be included. The lepton Lagrangian density reads
\begin{align}
\label{eqf2}
\mathcal{L}_l = \sum_{l=e,\mu}\bar{\psi}_l\left(i\gamma^\mu\partial_\mu-m_l\right)\psi_l.
\end{align}
The total Lagrangian density of the model can then be expressed as
\begin{align}
\label{eq54}
\mathcal{L}=\mathcal{L}_{B}+\mathcal{L}_{M}+\mathcal{L}_{l}.
\end{align}

For cold, static, homogeneous, and isotropic neutron star matter, the meson field operators are replaced by their ground-state expectation values within the mean-field approximation. Only the temporal components of the vector meson fields are retained, while the $\rho$ meson field has only its third component in isospin space,
\begin{align}
\sigma &\rightarrow \sigma_0, \quad
\sigma^* \rightarrow \sigma_0^*, \quad
\omega^\mu \rightarrow \delta^\mu_0 \omega_0, 
\nonumber\\
\phi^\mu &\rightarrow \delta^\mu_0 \phi_0, \quad
\boldsymbol{\rho}^{\,\mu}
\rightarrow
\delta^\mu_0(0,0,\rho_{03}).
\end{align}
The effective mass of baryon $B$ is determined by
the scalar meson fields and takes the form
\begin{align}
M_B^*=M_B-g_{\sigma B}\sigma_0-g_{\sigma^*B}\sigma_0^*.
\end{align}
The chemical potential of baryon $B$ is
\begin{align}
\mu_B=\sqrt{k_{F,B}^2\!+\!M_B^{*2}}\!+\!g_{\omega B}\omega_0\!+\!g_{\rho B}I_{3B}\rho_{03}\!+\!g_{\phi B}\phi_0,
\end{align}
where $k_{F,B}$ and $I_{3B}$ denote the Fermi momentum and the third component of isospin of baryon $B$, respectively.

The self-consistent equations for the meson mean fields follow from the Euler-Lagrange equations:
\begin{subequations}
\begin{align}
m_\sigma^2\sigma_0+g_2\sigma_0^2+g_3\sigma_0^3&=\sum_B g_{\sigma B} n_B^s,
\\
m_{\sigma^*}^2\sigma_0^*&=\sum_Y g_{\sigma^*Y} n_Y^s,
\\
m_\omega^2\omega_0+2\Lambda_v g_\omega^2 g_\rho^2\omega_0\rho_{03}^2&=\sum_B g_{\omega B} n_B,
\\
m_\rho^2\rho_{03}+2\Lambda_v g_\omega^2 g_\rho^2\omega_0^2\rho_{03}&=\sum_B g_{\rho B} I_{3B} n_B,
\\
m_\phi^2\phi_0&=\sum_Y g_{\phi Y} n_Y,
\end{align}
\end{subequations}
where $n_B$ and $n_B^s$ denote the vector and scalar densities of baryon $B$, respectively,
\begin{align}
n_B&=\frac{k_{F,B}^3}{3\pi^2},
\\
n_B^s&=\frac{1}{\pi^2}\int_0^{k_{F,B}}\frac{k^2 M_B^*}{\sqrt{k^2+M_B^{*2}}}dk.
\end{align}
The mean-field equations are solved self-consistently together with the Fermi momenta of the particle species, subject to $\beta$ equilibrium and charge neutrality:
\begin{align}
&\mu_B = \mu_n-q_B\mu_e,
\quad
\mu_e=\mu_\mu,
\\
&\sum_B q_B n_B-n_e-n_\mu = 0.
\end{align}
where $q_B$ denotes the electric charge number of baryon $B$.
\begin{align}
\varepsilon_{\rm BM}&=\sum_B \frac{1}{\pi^2}\int_0^{k_{F,B}}k^2\sqrt{k^2+M_B^{*2}}{\rm d}k
\nonumber\\
&+\frac{1}{2}m_\sigma^2\sigma_0^2+\frac{1}{3}g_2\sigma_0^3+\frac{1}{4}g_3\sigma_0^4+\frac{1}{2}m_{\sigma^*}^2\sigma_0^{*2}
\nonumber\\
&\!+\!\frac{1}{2}m_\omega^2\omega_0^2\!+\!\frac{1}{2}m_\rho^2\rho_{03}^2\!+\!\frac{1}{2}m_\phi^2\phi_0^2+3\Lambda_v g_\omega^2 g_\rho^2\omega_0^2\rho_{03}^2
\nonumber\\
&+\sum_{l=e,\mu}\frac{1}{\pi^2}\int_0^{k_{F,l}}k^2\sqrt{k^2+m_l^2}{\rm d}k.
\end{align}
\begin{align}
P_{\rm BM}&=\sum_B\frac{1}{3\pi^2}\int_0^{k_{F,B}}
\frac{k^4}{\sqrt{k^2+M_B^{*2}}}dk
\nonumber\\
&-\frac{1}{2}m_\sigma^2\sigma_0^2-\frac{1}{3}g_2\sigma_0^3 -\frac{1}{4}g_3\sigma_0^4-\frac{1}{2}m_{\sigma^*}^2\sigma_0^{*2}
\nonumber\\
&\!+\!\frac{1}{2}m_\omega^2\omega_0^2 \!+\!\frac{1}{2}m_\rho^2\rho_{03}^2\! +\! \frac{1}{2}m_\phi^2\phi_0^2 \!+\!\Lambda_v g_\omega^2 g_\rho^2 \omega_0^2\rho_{03}^2
\nonumber\\
&+\sum_{l=e,\mu} \frac{1}{3\pi^2} \int_0^{k_{F,l}} \frac{k^4}{\sqrt{k^2+m_l^2}}{\rm d}k,
\end{align}
The nucleon-meson coupling constants and nonlinear coupling parameters are generally constrained by the saturation properties of nuclear matter~\cite{Dutra:2014RMF}. The nuclear matter saturation properties and the corresponding nucleon-meson coupling constants adopted in this work are listed in Table~\ref{tab1}.
\begin{table*}[t]
\centering
\caption{RMF parameters for nucleonic matter adopted in this work and the corresponding nuclear matter saturation properties. Here, $m_\sigma$, $m_\omega$, and $m_\rho$ denote the masses of the $\sigma$, $\omega$, and $\rho$ mesons, respectively; $n_0$, $-E/A$, $m_N^*/m_N$, $K_0$, $E_{\rm sym}$, and $L$ represent the saturation density, binding energy per nucleon, nucleon effective mass ratio, incompressibility, symmetry energy, and its slope parameter, respectively. The parameters $g_\sigma$, $g_\omega$, and $g_\rho$ are the nucleon-meson coupling constants, $g_2$ and $g_3$ characterize the nonlinear self-interactions of the $\sigma$ field, and $\Lambda_v$ denotes the $\omega$-$\rho$ mixed coupling parameter.}
\label{tab1}
\renewcommand{\arraystretch}{1.}
\begin{tabular*}{\textwidth}{@{\extracolsep{\fill}}cccccccc@{}}
\toprule
$m_\sigma~(\mathrm{MeV})$
&
$m_\omega~(\mathrm{MeV})$
&
$m_\rho~(\mathrm{MeV})$
&
$n_0~(\mathrm{fm}^{-3})$
&
$-E/A~(\mathrm{MeV})$
&
$m_N^*/m_N$
&
$K_0~(\mathrm{MeV})$
&
$E_{\rm sym}~(\mathrm{MeV})$
\\

491.5
&
782.5
&
763
&
0.155
&
16.4
&
0.61
&
270
&
32.5
\\

\hline

\multicolumn{8}{c}{
\begin{tabular*}{0.96\textwidth}{@{\extracolsep{\fill}}ccccccc@{}}
$L~(\mathrm{MeV})$
&
$g_\sigma$
&
$g_\omega$
&
$g_\rho$
&
$g_2~(\mathrm{fm}^{-1})$
&
$g_3$
&
$\Lambda_v$
\\

60
&
9.53741
&
12.2827
&
10.4828
&
9.7508
&
$-24.525$
&
0.0286323
\end{tabular*}
}
\\
\bottomrule
\end{tabular*}
\end{table*}

\subsection{hyperon-meson coupling}
\label{sec2-2}
Within the RMF framework described above, the nucleon-meson coupling constants are fixed by the adopted parameter set. The inclusion of hyperonic degrees of freedom requires the additional determination of the coupling strengths between hyperons and meson fields. However, the hyperon-meson couplings remain subject to considerable uncertainties due to the limited experimental and theoretical constraints on hyperon interactions. Variations in these couplings affect the onset densities of hyperons and the particle composition of dense matter, thereby modifying its EOS. To characterize the hyperon-meson coupling strengths, we introduce the relative coupling constants
\begin{align}
x_{\sigma Y} = \frac{g_{\sigma Y}}{g_{\sigma N}}, \quad
x_{\omega Y} = \frac{g_{\omega Y}}{g_{\omega N}}, \quad
x_{\rho Y} = \frac{g_{\rho Y}}{g_{\rho N}},
\end{align}
where $g_{\sigma N}$, $g_{\omega N}$, and $g_{\rho N}$ denote the coupling constants of nucleons to the $\sigma$, $\omega$, and $\rho$ mesons, respectively, while $g_{\sigma Y}$, $g_{\omega Y}$, and $g_{\rho Y}$ represent the corresponding hyperon-meson coupling constants.

The couplings of hyperons to the vector mesons $\omega$ and $\rho$ are determined using SU(6) symmetry relations~\cite{Dutra:2014RMF}, whereas their couplings to the scalar meson $\sigma$ are determined from the single-particle potentials of hyperons in nuclear matter,
\begin{align}
U_Y^{(N)}
=
g_{\omega Y}\omega_0
-g_{\sigma Y}\sigma_0,
\end{align}
where $U_Y^{(N)}$ denotes the single-particle potential of hyperon $Y$ in symmetric nuclear matter. We adopt the commonly used empirical values $U_\Lambda^{(N)}=-28~\mathrm{MeV}$, $U_\Sigma^{(N)}=+30~\mathrm{MeV}$, and $U_\Xi^{(N)}=-18~\mathrm{MeV}$ ~\cite{Schaffner:1996Hyperon,Gal:2016Strangeness}. The resulting coupling parameters are listed in Table~\ref{tab2}.
\begin{table}[htbp]
\centering
\caption{Relative coupling constants between hyperons and the $\sigma$, $\omega$, and $\rho$ mesons.}
\label{tab2}
\setlength{\tabcolsep}{14pt}
\begin{tabular}{c c c c}
\toprule
 & $x_{\sigma Y}$ & $x_{\omega Y}$ & $x_{\rho Y}$ \\
\midrule
$\Lambda$ & 0.6106 & 0.6667 & 0 \\
$\Sigma$  & 0.4523 & 0.6667 & 2 \\
$\Xi$     & 0.3163 & 0.3333 & 1 \\
\bottomrule
\end{tabular}
\end{table}

To describe hyperon-hyperon interactions, we further introduce the strange mesons $\sigma^*$ and $\phi$, which are scalar and vector mesons, respectively. The $\sigma^*$ meson primarily provides an attractive interaction between hyperons, whereas the $\phi$ meson contributes a repulsive interaction~\cite{Schaffner:1996Hyperon,Schaffner:1994Strange}. Their relative coupling strengths are defined as $x_{\sigma^*Y}=g_{\sigma^*Y}/g_{\sigma N}$ and $x_{\phi Y}=g_{\phi Y}/g_{\omega N}$.

For the scalar strange meson $\sigma^*$, we consider three different hyperon coupling schemes:\\
(i) The coupling constants are determined using SU(6) symmetry relations~\cite{Schaffner:1996Hyperon,SchaffnerBielich2000,Sharma2026}, with
\begin{equation}
2g_{\sigma^*\Lambda}
=
2g_{\sigma^*\Sigma}
=
g_{\sigma^*\Xi}
=
\frac{2\sqrt{2}}{3}g_{\sigma N}.
\end{equation}
(ii) The coupling parameters are determined from the
hyperon-hyperon potential depths, with
\begin{equation}
U_{\Sigma}^{(\Xi)}
\simeq U_{\Lambda}^{(\Xi)}
\simeq U_{\Xi}^{(\Xi)}
\simeq 2U_{\Lambda}^{(\Lambda)}
\simeq 2U_{\Sigma}^{(\Lambda)}
\simeq 40~\mathrm{MeV},
\end{equation}
which yields the relative coupling constants
$g_{\sigma^*\Lambda}/g_{\sigma N}
=g_{\sigma^*\Sigma}/g_{\sigma N}=0.69$
and $g_{\sigma^*\Xi}/g_{\sigma N}=1.25$
\cite{Schaffner1993,Hong2018}.\\
(iii) The absolute coupling constants are adopted from Bednarek et al.~\cite{Bednarek:2017Hyperon}, with
\begin{equation}
g_{\sigma^\ast\Lambda}
=
g_{\sigma^\ast\Sigma}
=5.626,
\qquad
g_{\sigma^\ast\Xi}=11.623.
\end{equation}

For the vector strange meson $\phi$,
we consider two different hyperon coupling schemes:\\(i) The coupling constants are determined using SU(3) symmetry relations~\cite{Lopes2014}, with $\alpha_v=0.5$
\begin{equation}
\begin{aligned}
\frac{g_{\phi\Lambda}}{g_{\omega N}}
&=\frac{\sqrt{2}(2\alpha_v-5)}{5+4\alpha_v},\\
\frac{g_{\phi\Sigma}}{g_{\omega N}}
&=-\frac{\sqrt{2}(2\alpha_v+1)}{5+4\alpha_v},\\
\frac{g_{\phi\Xi}}{g_{\omega N}}
&=-\frac{\sqrt{2}(2\alpha_v+4)}{5+4\alpha_v}.
\end{aligned}
\label{eq:phi_SU3}
\end{equation}
(ii) The coupling constants are determined using SU(6) spin-flavor symmetry~\cite{Schaffner:1996Hyperon},
\begin{equation}
2g_{\phi\Lambda}
=
2g_{\phi\Sigma}
=
g_{\phi\Xi}
=
-\frac{2\sqrt{2}}{3}g_{\omega N}.
\end{equation}

By combining the three $\sigma^*$-hyperon coupling schemes with the two $\phi$-hyperon coupling schemes, we construct six EOSs, denoted by EOS1-EOS6. The six coupling schemes share the same nucleonic RMF model and differ only in the strange meson-hyperon coupling parameters, thereby representing different hyperon-hyperon interactions. The corresponding coupling constants are summarized in Table~\ref{tab3}.

\begin{table}[htbp]
\centering
\caption{Relative $\sigma^*$-hyperon and $\phi$-hyperon coupling constants for the six EOSs considered in this work.}
\label{tab3}
\setlength{\tabcolsep}{5pt}
\begin{tabular}{c c c c c c c}
\toprule
Case & $x_{\sigma^*\Lambda}$ & $x_{\sigma^*\Sigma}$ & $x_{\sigma^*\Xi}$
& $x_{\phi^\Lambda}$ & $x_{\phi^\Sigma}$ & $x_{\phi^\Xi}$\\
\midrule
EOS1 & 0.69  & 0.69  & 1.25 & -0.808  & -0.404   & -1.01  \\
EOS2 & 0.590  & 0.590  & 1.219 & -0.808  & -0.404   & -1.01 \\
EOS3 & 0.69  & 0.69  & 1.25 & -0.471 & -0.471 & -0.942  \\
EOS4 & 0.590  & 0.590  & 1.219 & -0.471 & -0.471 & -0.942 \\
EOS5 & 0.471 & 0.471 & 0.942 & -0.808  & -0.404   & -1.01  \\
EOS6 & 0.471 & 0.471 & 0.942 & -0.471 & -0.471 & -0.942 \\
\bottomrule
\end{tabular}
\end{table}

\subsection{Dark matter model}
\label{sec2-3}
For the dark matter component, we consider an interaction between dark matter and nucleonic-hyperonic matter mediated by Higgs boson exchange. Since this interaction is analogous to meson exchange between baryons, the system can be described by the single-fluid Tolman-Oppenheimer-Volkoff (TOV) equations, assuming that dark matter and baryonic matter are uniformly mixed within the neutron star. The dark matter candidate is taken to be a fermion $\chi$ with mass $m_\chi$. The corresponding Lagrangian density can be written as~\cite{Panotopoulos:2017Higgs,Das:2021Fmode}
\begin{align}
\mathcal{L}_{\rm DM}
&=
\bar{\chi}\left(i\gamma^\mu\partial_\mu-m_\chi+y_\chi h\right)\chi
+\frac{1}{2}\partial_\mu h\partial^\mu h
\nonumber\\
&-\frac{1}{2}m_h^2h^2+\sum_B \frac{f_B M_B}{v}\bar{\psi}_B h\psi_B.
\end{align}
Here, $h$ and $m_h$ denote the Higgs field and its mass, respectively, while $y_\chi$ is the dark matter-Higgs Yukawa coupling constant. The mass and Dirac field of baryon $B$ are represented by $M_B$ and $\psi_B$, respectively. The quantity $f_B M_B/v$ describes the effective Higgs-baryon Yukawa coupling strength, where $f_B$ is the coupling factor and $v$ is the Higgs vacuum expectation value. In this work, we adopt $m_h=125~\mathrm{GeV}$, $y_\chi=0.07$, $f_B=0.3$, and $v=246~\mathrm{GeV}$, with a dark matter particle mass of $m_\chi=200~\mathrm{GeV}$~\cite{Panotopoulos:2017Higgs,Hong:2024MixedDM,Murakami2001}.

Within the mean-field approximation, the Higgs field is replaced by its uniform ground-state expectation value $h_0$, which modifies the effective masses of dark matter and baryons through scalar couplings. The corresponding expressions are given by
\begin{align}
m_\chi^* &= m_\chi-y_\chi h_0, \\
M_B^* &= M_B-g_{\sigma B}\sigma_0-g_{\sigma^*B}\sigma_0^*-\frac{f_BM_B}{v}h_0 .
\end{align}

In this work, dark matter is treated as a zero-temperature degenerate Fermi gas, with its energy density and pressure given by
\begin{align}
\epsilon_{\rm DM}
&=
\frac{1}{\pi^2}
\int_0^{k_F^\chi}
k^2\sqrt{k^2+\left(m_\chi^*\right)^2}\,dk
+\frac{1}{2}m_h^2h_0^2,
\\
P_{\rm DM}
&=
\frac{1}{3\pi^2}
\int_0^{k_F^\chi}
\frac{k^4}{\sqrt{k^2+\left(m_\chi^*\right)^2}}\,dk
-\frac{1}{2}m_h^2h_0^2.
\end{align}
where $k_F^\chi$ denotes the dark matter Fermi momentum.

The total energy density and pressure of neutron star matter, including both baryonic and dark matter contributions, can be expressed as
\begin{align}
\epsilon_{\rm tot}
&=
\epsilon_{\rm BM}
+
\epsilon_{\rm DM},
\\
P_{\rm tot}
&=
P_{\rm BM}
+
P_{\rm DM},
\end{align}

For a given hyperon-meson coupling scheme and dark matter Fermi momentum $k_F^\chi$, the corresponding EOS of dark-matter-admixed hyperonic matter, $P_{\rm tot}=P_{\rm tot}(\epsilon_{\rm tot})$, is obtained by solving the mean-field equations self-consistently subject to $\beta$ equilibrium and charge neutrality. In the low-density crust region, the outer crust is described by the Baym-Pethick-Sutherland (BPS) EOS ~\cite{Baym:1971BPS}, while the inner crust is modeled using a polytropic EOS, $P=a+b\epsilon^{4/3}$~\cite{Carriere:2003Crust,GonzalezBoquera:2018Crust}, which is matched to the hyperon-dark matter EOS of the core at the crust-core transition.

\subsection{Stellar structure equations}
\label{sec2-4}
Once the complete EOS of dark-matter-admixed hyperonic matter is obtained, the macroscopic structure of the corresponding compact stars can be investigated. For static, spherically symmetric neutron stars in hydrostatic equilibrium, the stellar structure is governed by the Tolman-Oppenheimer-Volkoff (TOV) equations~\cite{Oppenheimer:1939TOV,Tolman:1939TOV}
\begin{align}
\frac{{\rm d}P_{\rm tot}}{{\rm d}r}&=-\frac{\left(\epsilon_{\rm tot}+P_{\rm tot}\right)\left[m(r)+4\pi r^3P_{\rm tot}\right]}{r\left[r-2m(r)\right]},
\\
      \frac{{\rm d}m(r)}{{\rm d}r}&=4\pi r^2\epsilon_{\rm tot}.
\end{align}
\section{Multimessenger constraints on dark matter admixed hyperon stars}
\label{sec3}
\subsection{Equations of state and stellar structure of hyperonic matter}
\label{sec3-1}
Before introducing dark matter degrees of freedom, we first investigate the effects of uncertainties in hyperon interactions on neutron star properties. We adopt EOS1-EOS6, which are constructed using the same nucleon-meson parameter set within the RMF framework but employ different hyperon-meson coupling schemes. We compare their equations of state and mass-radius relations, and further examine whether variations in the hyperon-meson couplings alone can simultaneously support massive neutron stars and reproduce the low-mass compact configuration associated with HESS J1731-347.
\begin{figure}[htbp]
\centering
\includegraphics[width=1.0\linewidth]{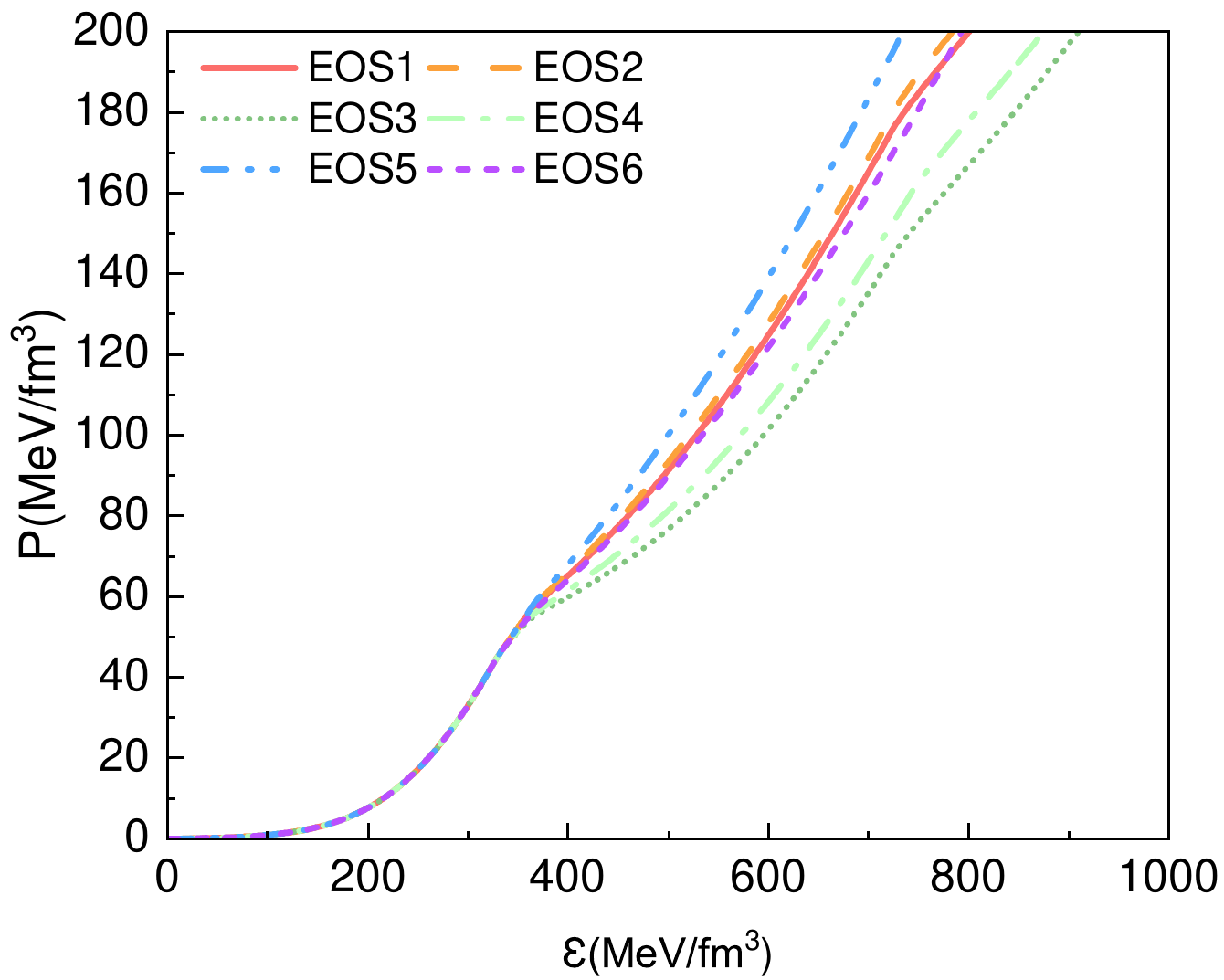}
\caption{Pressure-energy density relations of hyperonic matter for the six hyperon-meson coupling schemes.}
\label{fig1}
\end{figure}

Figure~\ref{fig1} shows the pressure-energy density relations for EOS1-EOS6. At low energy densities, the six EOSs nearly overlap. As the energy density increases, the curves gradually separate, exhibiting noticeable differences at high densities. Among them, EOS5 is relatively stiff, whereas EOS3 and EOS4 are comparatively soft, while EOS1, EOS2, and EOS6 exhibit similar high-density behavior. Since the models employ the same nucleon-meson coupling parameters, these differences arise primarily from the contributions of the strange mesons $\sigma^*$ and $\phi$, which govern the attractive and repulsive interactions between hyperons, respectively. Overall, the uncertainties in hyperon interactions are mainly reflected in the high-density behavior of the EOS, whereas their effects at lower densities are relatively limited. Whether these differences in the high-density EOS translate into changes in the macroscopic stellar structure requires further examination of the corresponding mass-radius relations.
\begin{figure}
\centering
\includegraphics[width=1.0\linewidth]{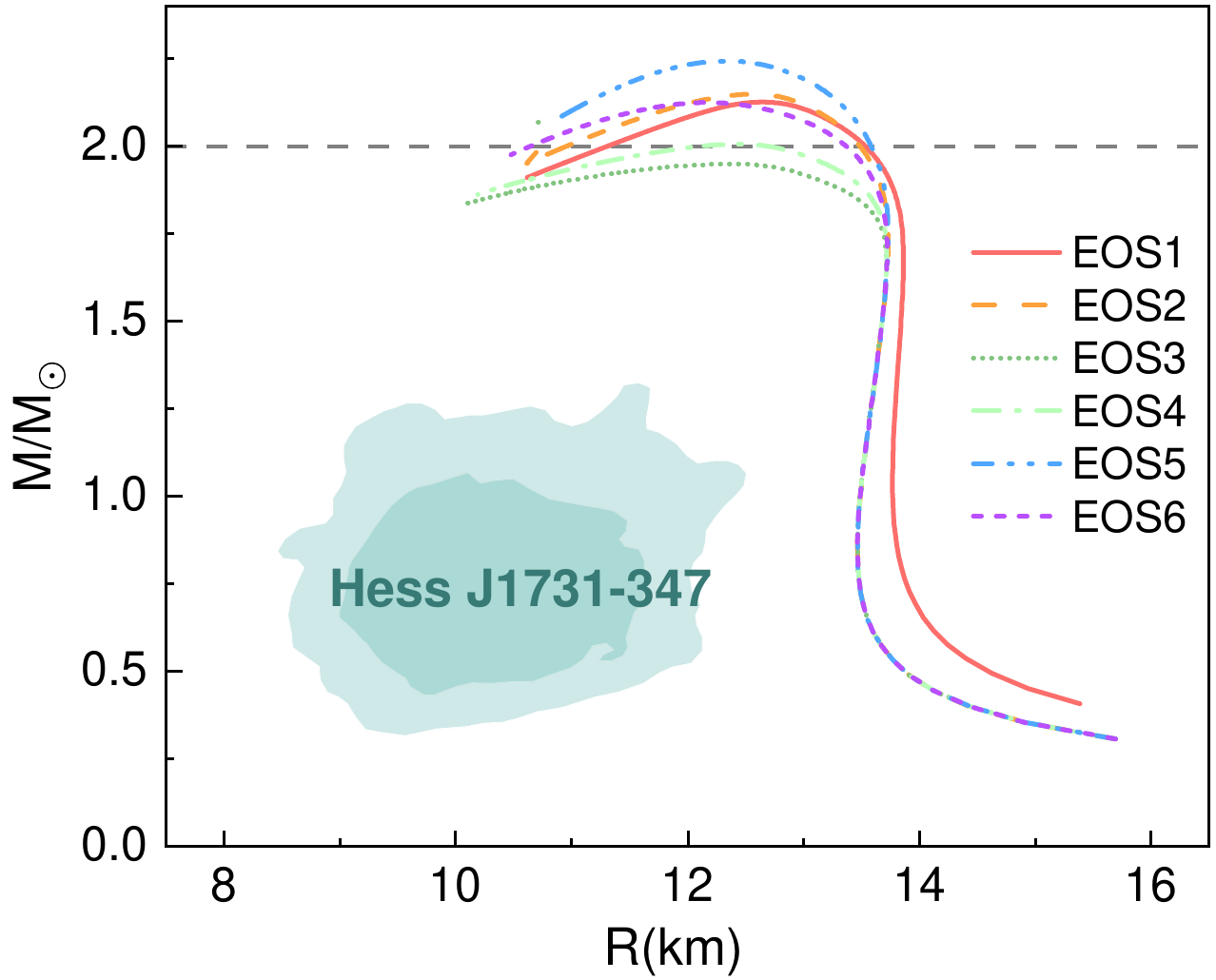}
\caption{Mass-radius relations of neutron stars for the six hyperon-meson coupling schemes.  The dark and light shaded regions correspond to the 68.3\% and 95.4\% credible regions for HESS J1731$-$347, respectively.}
\label{fig2}
\end{figure}

To further investigate the effects of differences in the EOS on the macroscopic stellar structure, we present the mass-radius relations for EOS1-EOS6 in Fig.~\ref{fig2}. The results show that the effects of different hyperon-meson coupling schemes on stellar structure exhibit a pronounced mass dependence. In the high-mass region, the curves gradually diverge, and the maximum mass $M_{\max}$ is particularly sensitive to hyperon interactions. The six models yield maximum masses ranging from approximately $1.95M_\odot$ to $2.24M_\odot$, with EOS5 giving the largest value and EOS3 the smallest. Taken together, Figs.~\ref{fig1} and \ref{fig2} indicate that a stiffer EOS at high densities generally supports a larger maximum neutron star mass. This is because, although $M_{\max}$ is determined by the full EOS, it is particularly sensitive to the EOS at high densities.

In contrast, at low stellar masses, the mass-radius curves for EOS2-EOS6 lie close to one another, whereas EOS1 yields larger stellar radii at a given mass. In this mass range, HESS J1731-347 provides an important observational constraint. Doroshenko et al.~\cite{Doroshenko:2022HESS}, using X-ray spectral modeling combined with distance estimates from Gaia observations, inferred a mass of $M=0.77_{-0.17}^{+0.20}M_\odot$ and a radius of $R=10.4_{-0.78}^{+0.86}~\mathrm{km}$ (68.3\% credible level) for the central compact object in the supernova remnant HESS J1731-347. Figure~\ref{fig2} also shows the 68.3\% and 95.4\% credible regions of its joint mass-radius distribution. However, none of the mass-radius curves for the six models enter the 95.4\% credible region of HESS J1731-347. This is mainly because low-mass neutron stars have relatively low central densities, where hyperonic degrees of freedom have a limited effect on the EOS. Consequently, varying the hyperon-meson coupling parameters alone cannot significantly reduce the stellar radius.

These results suggest that, within the hyperonic models considered in this work, variations in hyperonic degrees of freedom and their interactions alone are insufficient to reproduce the low-mass, highly compact configuration associated with HESS J1731-347, indicating the need to consider the effects of additional physical degrees of freedom.
\subsection{HESS J1731$-$347 constraint}
\label{sec3-2}
On this basis, we introduce dark matter degrees of freedom and use the dark matter Fermi momentum $k_F^\chi$ to characterize the dark matter component. For fermionic dark matter at zero temperature, the Fermi momentum is determined by its number density. A dark matter number density of approximately $10^{-3}$ times the nuclear saturation density corresponds to $k_F^\chi\simeq0.03~\mathrm{GeV}$. Previous studies have typically investigated variations in the dark matter Fermi momentum on the order of $10^{-2}~\mathrm{GeV}$. For example, Das et al.~\cite{Das2020} examined the effects of varying the dark matter Fermi momentum on the macroscopic properties of neutron stars for $k_F^{\rm DM}\lesssim0.06~\mathrm{GeV}$, while Kumar et al.~\cite{Kumar2022} investigated the thermal evolution of dark-matter-admixed neutron stars over the range $0.02\leq k_F^{\rm DM}\leq0.06~\mathrm{GeV}$. Motivated by these studies, we consider stellar configurations for EOS1-EOS6 with $0.020\leq k_F^\chi\leq0.060~\mathrm{GeV}$ and obtain the corresponding mass-radius relations by solving the TOV equations. Comparison with the mass-radius constraints from HESS J1731-347 then allows us to determine the permitted ranges of $k_F^\chi$ for different hyperon-meson coupling schemes.
\begin{figure*}[htbp]
\centering
\includegraphics[width=1.0\linewidth]{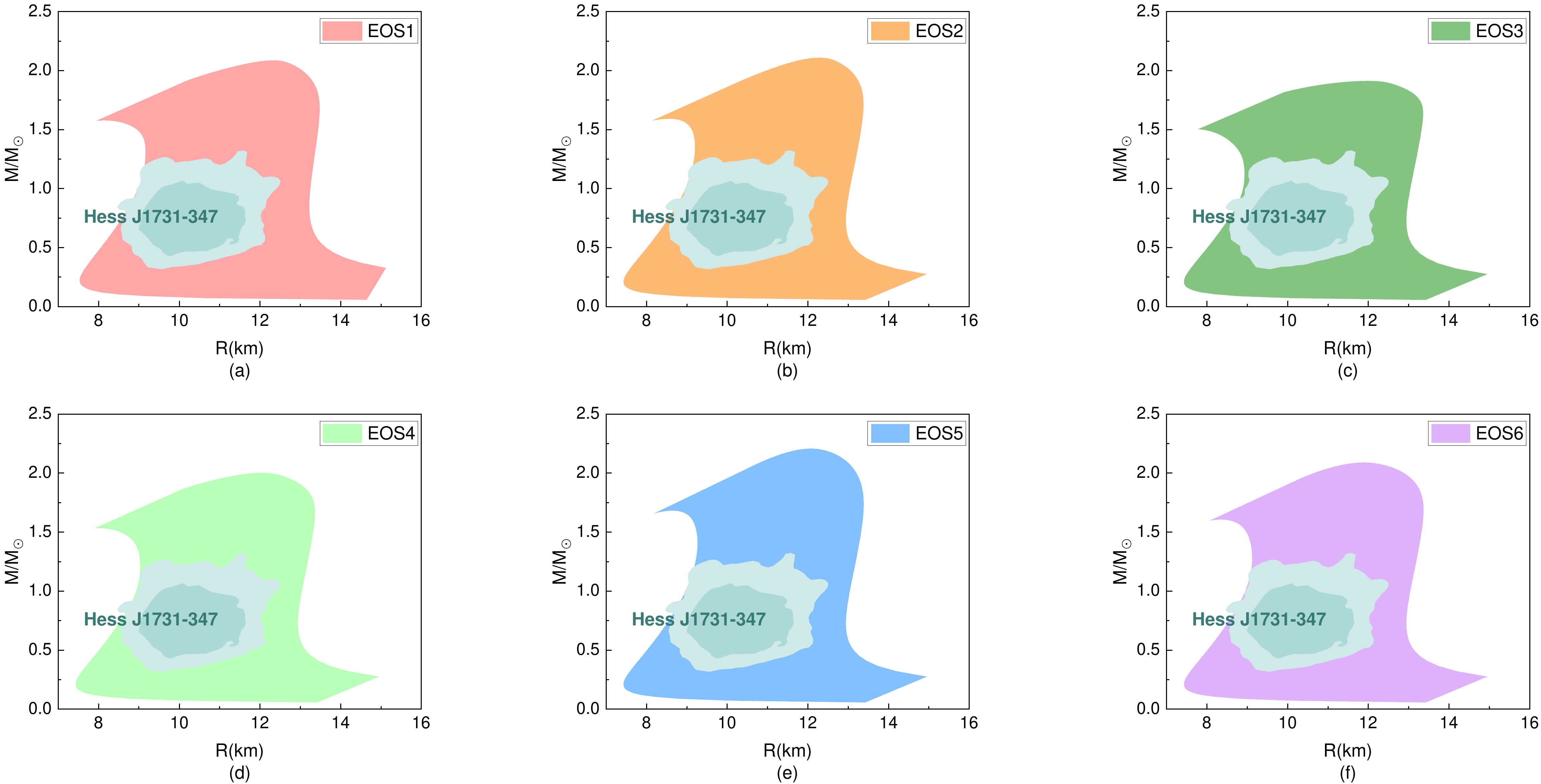}
\caption{Mass-radius relations for different hyperon-meson coupling schemes
with the dark matter Fermi momentum in the range
$0.020\leq k_F^\chi\leq 0.060~\mathrm{GeV}$.
The dark and light shaded regions correspond to the
68.3\% and 95.4\% credible regions for
HESS J1731$-$347, respectively.}
\label{fig3}
\end{figure*}

Figure~\ref{fig3} shows the mass-radius relations for EOS1-EOS6 at different values of the dark matter Fermi momentum $k_F^\chi$. As $k_F^\chi$ increases, the maximum mass $M_{\max}$ supported by each model gradually decreases, indicating a significant effect of the dark matter component on the maximum neutron star mass. Meanwhile, in the low-mass region, a larger $k_F^\chi$ corresponds to a smaller stellar radius at a given mass. This allows all six hyperonic models, which fail to satisfy the observational constraints from HESS J1731-347 in the absence of dark matter, to become consistent with its mass-radius observations for appropriate ranges of $k_F^\chi$.

Since the radii of low-mass neutron stars are relatively sensitive to $k_F^\chi$, the mass-radius measurements of HESS J1731-347 can place meaningful constraints on the dark matter Fermi momentum. As shown in Fig.~\ref{fig3} and Table~\ref{tab4}, at the 95.4\% credible level, the allowed ranges are $0.029\leq k_F^\chi\leq0.060~\mathrm{GeV}$ for EOS1 and $0.028\leq k_F^\chi\leq0.060~\mathrm{GeV}$ for EOS2, while the allowed range for each of EOS3-EOS6 is $0.027\leq k_F^\chi\leq0.060~\mathrm{GeV}$. When the 68.3\% credible region is considered, the allowed range narrows to $0.034\leq k_F^\chi\leq0.056~\mathrm{GeV}$ for EOS2 and $0.033\leq k_F^\chi\leq0.056~\mathrm{GeV}$ for the other five models. Compared with the initial scan range, the observational constraints from HESS J1731-347 exclude certain values of the dark matter Fermi momentum, with the 68.3\% credible region yielding narrower allowed ranges.

Although HESS J1731-347 provides effective constraints on the dark matter Fermi momentum, six hyperon-meson coupling schemes can satisfy its mass-radius constraints at both credible levels for suitable values of $k_F^\chi$, with very similar allowed ranges among the different schemes. This suggests that the observations of HESS J1731-347 alone are not sufficient to distinguish effectively among the different hyperon-meson coupling schemes. The main reason is that hyperonic degrees of freedom have a relatively limited influence on the EOS and stellar structure of low-mass neutron stars. Therefore, to further distinguish among these coupling schemes, it is necessary to incorporate observations of neutron stars with typical and high masses to place additional constraints on the high-density behavior of the EOS.
\subsection{NICER and massive-pulsar constraints}
\label{sec 3-3}
To further investigate the stellar structures associated with different hyperon-meson coupling schemes, we incorporate the mass-radius constraints inferred from NICER observations of PSR J0030+0451 and PSR J0740+6620, in addition to those from HESS J1731-347, and require each model to satisfy $M_{\max}\geq 2M_\odot$. This allows us to examine whether the models can simultaneously satisfy neutron star observations across different mass ranges.

For PSR J0030+0451, Riley et al.~\cite{Riley2019} inferred a mass of $M=1.34^{+0.15}_{-0.16}M_\odot$ and a radius of $R=12.71^{+1.14}_{-1.19}~\mathrm{km}$ (68.3\% credible level) based on X-ray pulse-profile observations with NICER, while Miller et al.~\cite{Miller:2019J0030} obtained $M=1.44^{+0.15}_{-0.14}M_\odot$ and $R=13.02^{+1.24}_{-1.06}~\mathrm{km}$ (68.3\% credible level). For the massive pulsar PSR J0740+6620, Riley et al.~\cite{Riley:2021J0740} combined X-ray observations from NICER and XMM-Newton with radio-timing mass constraints, obtaining $M=2.072^{+0.067}_{-0.066}M_\odot$ and $R=12.39^{+1.30}_{-0.98}~\mathrm{km}$ (68.3\% credible level). Miller et al.~\cite{Miller:2021J0740}, using a radio-timing mass constraint of $M=2.08\pm0.07M_\odot$, inferred a radius of $R=13.7^{+2.6}_{-1.5}~\mathrm{km}$ (68.3\% credible level). In addition, Salmi et al.~\cite{Salmi:2024J0740} further updated the radius estimate for PSR J0740+6620 using a longer NICER observational data set. Compared with the low-mass neutron star constraints provided by HESS J1731-347, PSR J0030+0451 and PSR J0740+6620 provide constraints on the structures of neutron stars with typical masses and massive neutron stars, respectively. These observations thus allow further tests of the EOS at higher densities, where the differences among the hyperon-meson coupling schemes become more pronounced.

\begin{figure*}[htbp]
\centering
\includegraphics[width=1.0\linewidth]{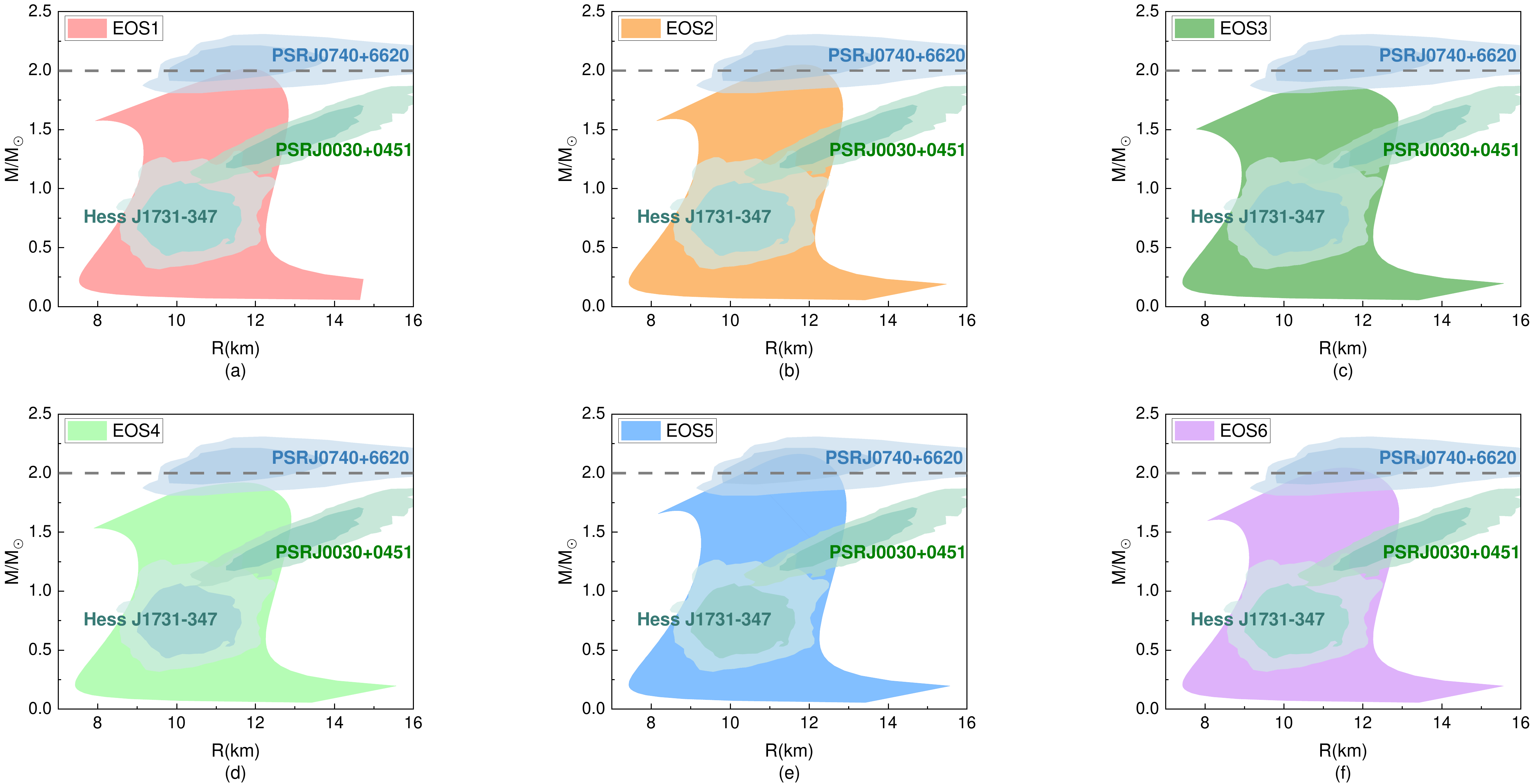}
\caption{Mass-radius relations under astronomical observational constraints for the six hyperon-meson coupling schemes. The shaded regions represent the mass-radius constraints
for three neutron stars: PSR J0740+6620,
PSR J0030+0451, and HESS J1731$-$347
(at the 68.3\% and 95.4\% credible levels). Panels (a)-(f) correspond to EOS1-EOS6, respectively.}
\label{fig4}
\end{figure*}

In this work, we adopt the joint mass-radius posterior distributions for PSR J0030+0451 and PSR J0740+6620 obtained by Miller et al.~\cite{Miller:2019J0030,Miller:2021J0740}. Their 68.3\% and 95.4\% credible regions are shown in Fig.~\ref{fig4} to constrain the hyperon-meson coupling schemes and dark matter parameters. The results show that the mass-radius observations of the two pulsars place different constraints on the six EOSs. Consequently, the allowed parameter spaces vary significantly among the models. At the 95.4\% credible level, the allowed ranges of $k_F^\chi$ for EOS1 and EOS2 satisfying the observational constraints from PSR J0030+0451 are $0.029\leq k_F^\chi\leq0.049~\mathrm{GeV}$ and $0.028\leq k_F^\chi\leq0.049~\mathrm{GeV}$, respectively. For each of EOS3-EOS6, the allowed range is $0.027\leq k_F^\chi\leq0.049~\mathrm{GeV}$. At the 68.3\% credible level, the six models can still satisfy the observational constraints within certain parameter ranges, although their allowed ranges become narrower. These results show that the mass-radius observations of PSR J0030+0451 can constrain the dark matter Fermi momentum but are insufficient to effectively distinguish among the different hyperon-meson coupling schemes.

In contrast, the observations of PSR J0740+6620 have greater discriminating power among different hyperon-meson coupling schemes. At the 95.4\% credible level, the allowed ranges of $k_F^\chi$ for EOS1 and EOS2 satisfying the mass-radius constraints of PSR J0740+6620 are $0.029\leq k_F^\chi\leq0.044~\mathrm{GeV}$ and $0.028\leq k_F^\chi\leq0.046~\mathrm{GeV}$, respectively, while EOS5 and EOS6 have the same allowed range of $0.027\leq k_F^\chi\leq0.047~\mathrm{GeV}$. Notably, neither EOS3 nor EOS4 satisfies these constraints within the considered range of $k_F^\chi$, and both models are therefore excluded. When the 68.3\% credible region is considered, the allowed parameter ranges for the remaining four models become narrower. These results demonstrate that mass-radius observations of massive neutron stars can more effectively probe differences in the high-density EOS arising from different hyperon interactions.

In addition to the mass-radius constraints discussed above, we impose a lower limit of $M_{\max}\geq 2M_\odot$ on the maximum neutron star mass to examine whether the remaining four models can support massive neutron stars. As shown in Table~\ref{tab4}, the allowed ranges of the dark matter Fermi momentum are further reduced to $0.029\leq k_F^\chi\leq0.030~\mathrm{GeV}$ for EOS1, $0.028\leq k_F^\chi\leq0.032~\mathrm{GeV}$ for EOS2, $0.027\leq k_F^\chi\leq0.040~\mathrm{GeV}$ for EOS5, and $0.027\leq k_F^\chi\leq0.031~\mathrm{GeV}$ for EOS6. Although this additional constraint does not rule out any of the four models, it substantially reduces their allowed parameter space.

Overall, mass-radius observations of neutron stars with typical masses and massive neutron stars can effectively distinguish among different hyperon-meson coupling schemes and further constrain the dark matter parameter space. However, the remaining four EOSs still admit ranges of the dark matter Fermi momentum that satisfy the combined observational constraints, indicating that the degeneracy among the models has not been completely resolved. To further test the consistency of the remaining models with observations, we consider the tidal deformability constraints from GW170817 in the next section and investigate their impact on the dark matter parameter space.
\subsection{Tidal deformability constraint}
\label{sec3-4}
In binary neutron star systems, a neutron star undergoes tidal deformation in response to the tidal field of its companion. This tidal response leaves an observable imprint on the gravitational-wave signal during the inspiral phase, providing important information about the stellar interior and constraints on the EOS. The response of a neutron star to an external tidal field is commonly characterized by the dimensionless tidal deformability $\Lambda$~\cite{Flanagan:2008Tidal,Hinderer:2010Tidal}. For a static, spherically symmetric neutron star, $\Lambda$ is defined as
\begin{equation}
\Lambda=\frac{2}{3}k_2 C^{-5},
\end{equation}
where $C=M/R$ is the stellar compactness, $M$ and $R$ denote the gravitational mass and radius of the neutron star, respectively, and $k_2$ is the quadrupolar tidal Love number~\cite{Hinderer:2007mb}.

For a given static, spherically symmetric stellar configuration, the quadrupole tidal Love number $k_2$ is determined by the compactness $C$ and by the tidal perturbation variable at the stellar surface, $y_R\equiv y(R)$~\cite{Hinderer:2007mb,Damour:2009Tidal},
\begin{align}
k_2={}&\frac{8C^5}{5}(1-2C)^2
\left[2+2C(y_R-1)-y_R\right] 
\nonumber\\
&\times
\left\{
2C\left[6-3y_R+3C(5y_R-8)\right]+4C^3\right.
\nonumber\\
&\times \left[13-11y_R+C(3y_R-2)+2C^2(1+y_R)\right]
\nonumber\\
& +3(1-2C)^2\left[2-y_R+2C(y_R-1)\right]
\nonumber\\
&\left. \times \ln(1-2C)\right\}^{-1}.
\end{align}
$y(r)$ satisfies the static quadrupole tidal perturbation equation
\begin{equation}
r\frac{{\rm d}y}{{\rm d}r}+y^2+yF(r)+r^2Q(r)=0,
\end{equation}
where $F(r)$ and $Q(r)$ depend on the internal structure of the star and on its EOS. For the single-fluid hyperon-dark matter model considered here, the tidal Love number $k_2$ is determined jointly by the TOV equations and the tidal perturbation equation~\cite{Hinderer:2010Tidal,Postnikov:2010Tidal}.

\begin{figure*}[htbp]
\centering
\includegraphics[width=0.8\linewidth]{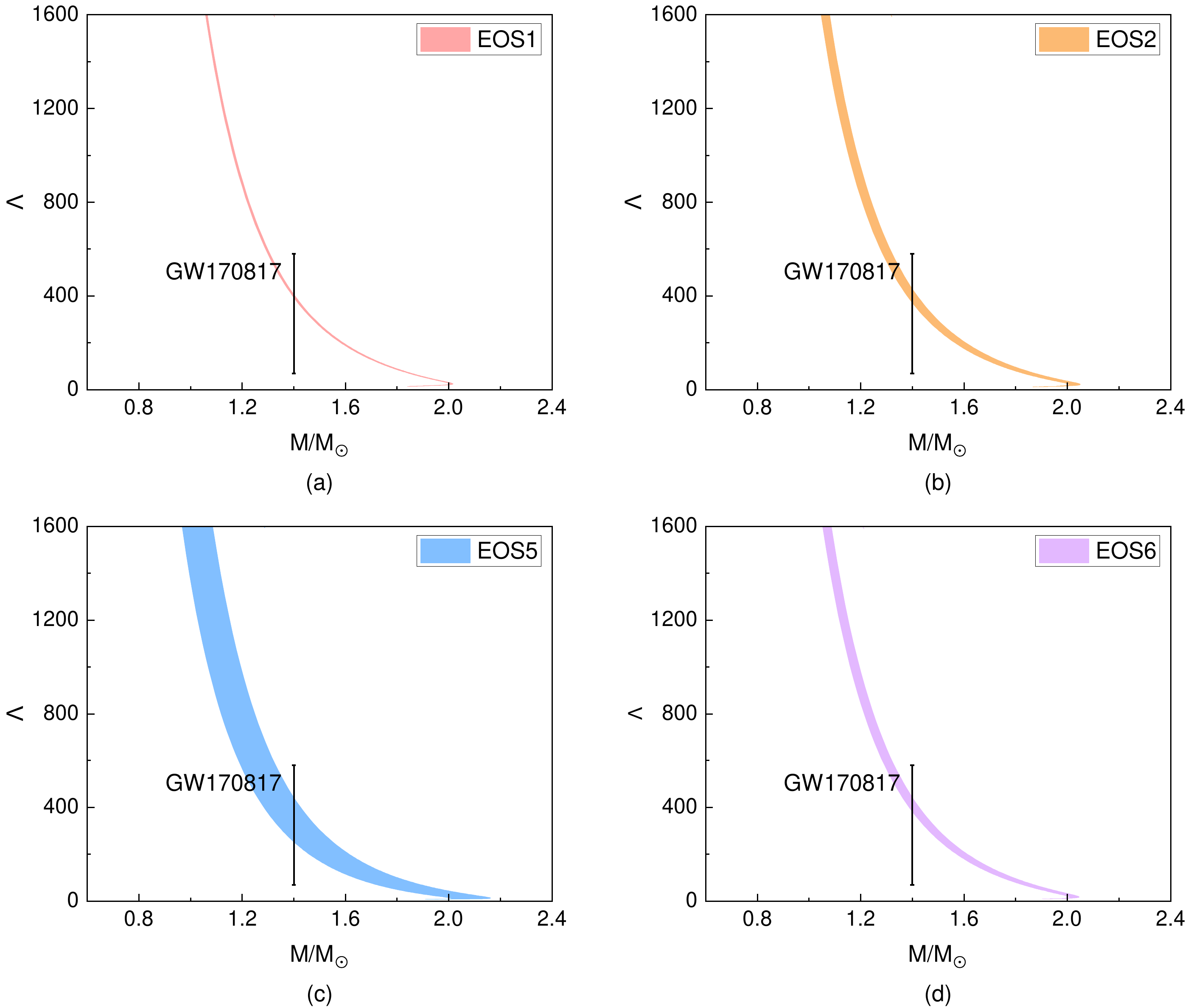}
\caption{The dimensionless tidal deformability $\Lambda$ as a function of neutron-star mass for EOS1, EOS2, EOS5, and EOS6. Panels (a)-(d) show the results corresponding to the dark matter Fermi momentum ranges allowed by the mass-radius constraints and the maximum-mass requirement. The black vertical lines indicate the 90\% credible interval of the tidal deformability at $1.4M_\odot$ inferred from GW170817.}
\label{fig5}
\end{figure*}

Gravitational-wave observations of the binary neutron star merger GW170817 have provided direct constraints on neutron star tidal deformability. Abbott et al.~\cite{Abbott:2018exr} obtained a dimensionless tidal deformability of $\Lambda_{1.4}=190^{+390}_{-120}$ for a canonical $1.4M_\odot$ neutron star at the 90\% credible level. To further investigate the tidal properties of the remaining models, we calculate the dimensionless tidal deformability $\Lambda$ as a function of neutron star mass $M$ for EOS1, EOS2, EOS5, and EOS6, as shown in Fig.~\ref{fig5}. We then compare the predicted values of $\Lambda_{1.4}$ for each model over its allowed range of $k_F^\chi$ with the tidal deformability constraint from GW170817 to assess their consistency with the observations.

As shown in Fig.~\ref{fig5}, the dimensionless tidal deformability $\Lambda$ exhibits a similar mass dependence for the different hyperon-meson coupling schemes. As the neutron star mass increases, $\Lambda$ decreases significantly. On the other hand, at a given stellar mass, $\Lambda$ gradually decreases with increasing $k_F^\chi$. Further examination of the tidal deformability at $M=1.4M_\odot$ shows that, within the allowed ranges of the dark matter Fermi momentum determined by the preceding mass-radius observational constraints and maximum-mass requirement, the values of $\Lambda_{1.4}$ for all four remaining models lie within the 90\% credible interval inferred from GW170817. Therefore, the tidal deformability observations do not further reduce their allowed parameter spaces.

To compare the constraints imposed by different astrophysical observations on the model parameter space, Table~\ref{tab4} summarizes the allowed ranges of the dark matter Fermi momentum for EOS1-EOS6 under different observational constraints. Specifically, the mass-radius observations of HESS J1731-347 primarily constrain the dark matter Fermi momentum, while the NICER observations further distinguish among different hyperon-meson coupling schemes. In particular, the mass-radius observations of PSR J0740+6620 rule out EOS3 and EOS4. The maximum-mass constraint further narrows the allowed parameter spaces of the remaining four models, whereas the tidal deformability observations from GW170817 impose no additional restrictions. In summary, these complementary astrophysical observations significantly reduce the allowed parameter space of the hyperon-dark matter models, although the degeneracy among different microscopic models has not been completely resolved.
\begin{table*}[htbp]
\centering
\caption{Allowed ranges of the dark-matter Fermi momentum $k_F^\chi$ for EOS1-EOS6 under different astrophysical observational constraints. The HESS J1731-347 and NICER mass-radius constraints are evaluated at the 95.4\% credible level, while the GW170817 tidal deformability constraint is imposed at the 90\% credible level.}
\label{tab4}
\begin{tabular}{c c c c c c c}
\toprule
Case &
\multicolumn{6}{c}{$k_F^\chi$ (GeV)} \\
\cmidrule(lr){2-7}
& Initial & HESS allowed & +NICER & $+M_{\max}\geq 2M_\odot$ & +GW170817 & Final \\
\midrule
EOS1 & 0.020-0.060 & 0.029-0.060 &0.029-0.044& 0.029-0.030 & 0.029-0.030& 0.029-0.030\\
EOS2 & 0.020-0.060 & 0.028-0.060 & 0.028-0.046
     & 0.028-0.032 & 0.028-0.032 & 0.028-0.032\\
EOS3 & 0.020-0.060 & 0.027-0.060 & Excluded 
     & - & -& - \\
EOS4 & 0.020-0.060 & 0.027-0.060 & Excluded 
     & - & -& -\\
EOS5 & 0.020-0.060 & 0.027-0.060 & 0.027-0.047
     &0.027-0.040 & 0.027-0.040& 0.027-0.040 \\
EOS6 & 0.020-0.060 & 0.027-0.060 &0.027-0.047
     &0.027-0.031 & 0.027-0.031 & 0.027-0.031\\
\bottomrule
\end{tabular}%
\end{table*}
\subsection{Nonradial $f$-mode oscillation frequencies}
\label{sec3-5}
Although current multimessenger observations have effectively constrained the parameter space of hyperon-dark matter models, several models with different hyperon interactions and dark matter contents remain consistent with the observational constraints. To investigate the differences in the dynamical properties of these remaining allowed models, we further study their nonradial $f$-mode oscillation frequencies. Nonradial oscillations of neutron stars are potential sources of gravitational waves, and their oscillation properties can reflect the stellar internal structure and matter composition. Among the various oscillation modes, the fundamental $f$-mode is considered one of the most promising nonradial modes for producing detectable gravitational-wave emission~\cite{Andersson:1997rn,Kokkotas:1999QNM}. Therefore, theoretical predictions of the $f$-mode frequencies for models satisfying current observational constraints can provide dynamical information complementary to that obtained from mass-radius relations and tidal deformability. Such predictions also provide a theoretical basis for distinguishing different hyperon-meson coupling schemes and dark matter components through future observations of stellar oscillations.

In calculating the $f$-mode oscillation frequencies, we adopt the relativistic Cowling approximation, in which perturbations of the spacetime metric during stellar oscillations are neglected, and only perturbations of the fluid variables are considered~\cite{Cowling:1941Oscillations}. Previous studies have shown that the $f$-mode frequencies obtained within the Cowling approximation deviate by approximately 20\% from those calculated using the full linearized equations of general relativity~\cite{Yoshida:1997Cowling,Rather:2025NonradialHyperons}. For a static, spherically symmetric background, the spacetime metric of the star can be written as
\begin{align}
{\rm d}s^2\!=\!\!-e^{2\Phi(r)}{\rm d}t^2\!+\!e^{2\lambda(r)}{\rm d}r^2\!+\!r^2\! \left({\rm d}\theta^2\!+\!\sin^2\theta{\rm d}\phi^2\right)\!,
\end{align}
where $\Phi(r)$ and $\lambda(r)$ are the metric functions, which satisfy
\begin{align}
e^{-2\lambda(r)}=1-\frac{2m(r)}{r}.
\end{align}

Within the Cowling approximation, nonradial oscillations can be described by two radial perturbation functions, $W(r)$ and $V(r)$, which satisfy the following equations~ \cite{Sotani:2011Fmode,RaneaSandoval:2018Cowling}:
\begin{align}
\frac{{\rm d}W(r)}{{\rm d}r}
=&\frac{{\rm d}\varepsilon}{{\rm d}P}
\left[
\omega^2 r^2 e^{\lambda(r)-2\Phi(r)}V(r)
+\frac{{\rm d}\Phi(r)}{{\rm d}r}W(r)
\right] 
\nonumber\\
&-l(l+1)e^{\lambda(r)}V(r),
\end{align}
and
\begin{align}
\frac{{\rm d}V(r)}{{\rm d}r}=2\frac{d\Phi(r)}{dr}V(r)-\frac{1}{r^2}e^{\lambda(r)}W(r).
\end{align}
Here, $\omega$ denotes the oscillation eigenfrequency, $l$ is the angular quantum number, and $\varepsilon$ and $P$ represent the total energy density and pressure of the single-fluid hyperon-dark matter system, respectively. The metric function $\Phi(r)$ satisfies the following hydrostatic equilibrium relation
\begin{align}
\frac{{\rm d}\Phi(r)}{{\rm d}r}
=
-\frac{1}{\varepsilon(r)+P(r)}
\frac{{\rm d}P(r)}{{\rm d}r}.
\end{align}

To ensure that the perturbation solutions remain regular at the stellar center, we impose the following boundary conditions near $r=0$~\cite{RaneaSandoval:2018Cowling}
\begin{align}
W(r)=Ar^{l+1},
\quad
V(r)=-\frac{A}{l}r^l,
\end{align}
where $A$ is an arbitrary normalization constant. At the stellar surface $r=R$, the Lagrangian pressure perturbation is required to vanish, yielding
\begin{align}
\omega^2 e^{\lambda(R)-2\Phi(R)}V(R)\!+\!\frac{1}{R^2}\!\left.\frac{{\rm d}\Phi(r)}{{\rm d}r}\right|_{r=R}
W(R)=0.
\end{align}
In this work, we focus on the most representative quadrupolar ($l=2$) fundamental $f$-mode and investigate how its oscillation frequency varies with neutron star mass, different hyperon-meson coupling schemes, and the dark matter Fermi momentum.
\begin{figure*}[htbp]
\centering
\includegraphics[width=0.8\linewidth]{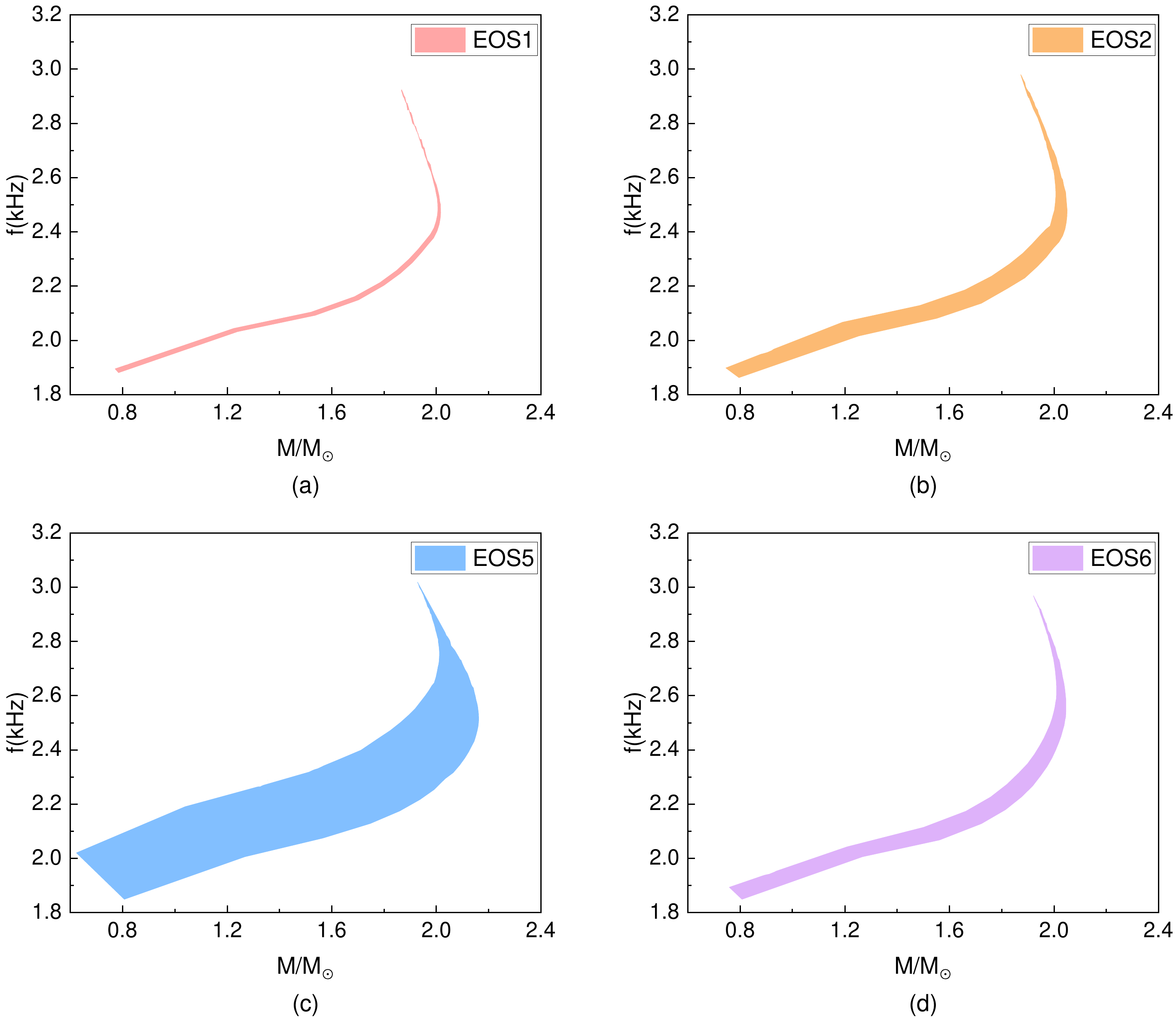}
\caption{The $f$-mode oscillation frequency as a function of neutron star mass for EOS1, EOS2, EOS5, and EOS6 after imposing the multimessenger constraints.}
\label{fig6}
\end{figure*}
Figure~\ref{fig6} shows the $f$-mode oscillation frequencies as a function of neutron star mass for the four models that satisfy the multimessenger observational constraints. The oscillation frequency generally increases with stellar mass, while the curve bends after reaching the maximum mass. To investigate the effects of different hyperon-meson coupling schemes, we fix the dark matter Fermi momentum at $k_F^\chi=0.030~\mathrm{GeV}$. At this value, the oscillation frequencies of the four models at $1.4M_\odot$ are all approximately $2.08~\mathrm{kHz}$, with a maximum difference of only $3.60~\mathrm{Hz}$. In contrast, at $2.0M_\odot$, the frequencies range from $2.33$ to $2.50~\mathrm{kHz}$, with a maximum difference of $172.65~\mathrm{Hz}$. This comparison suggests that the $f$-mode frequency can provide a useful probe of differences among hyperon-meson coupling schemes, particularly in massive neutron stars.

We further examine the effect of the dark matter Fermi momentum on the $f$-mode oscillation frequency. For all four models, the oscillation frequency increases with increasing dark matter Fermi momentum. Taking EOS5 as an example, as $k_F^\chi$ increases from $0.027$ to $0.040~\mathrm{GeV}$, the oscillation frequency of a $1.4M_\odot$ neutron star increases from $2.03650$ to $2.28608~\mathrm{kHz}$, corresponding to an increase of $249.58~\mathrm{Hz}$. At $2.0M_\odot$, the frequency increases from $2.26146$ to $2.67487~\mathrm{kHz}$, an increase of $413.41~\mathrm{Hz}$. These results indicate that, within the observationally allowed parameter range, the effect of the dark matter Fermi momentum on the $f$-mode oscillation frequency also exhibits a pronounced mass dependence, with a stronger effect in massive neutron stars.

Overall, variations in hyperon-meson coupling schemes and dark matter parameters can still produce noticeable differences in the $f$-mode oscillation frequencies, even for models satisfying current mass-radius and tidal deformability constraints. Although direct observational constraints on neutron star $f$-mode frequencies are not yet available, these theoretical results may provide a useful reference for future observations. The improved sensitivity of third-generation gravitational-wave detectors may enable the detection of neutron star $f$-mode signals, offering complementary information on the internal composition of dense matter and the effects of dark matter on neutron star structure.
\section{Summary}
\label{sec4}
In this work, we investigate the structure
and related observational properties of neutron stars containing both hyperonic degrees of freedom and fermionic dark matter within a single-fluid framework based on relativistic mean-field (RMF) theory. Given the considerable uncertainties in hyperon interactions at high densities, six different hyperon-meson coupling schemes are constructed within a common nucleon-meson framework. Fermionic dark matter interacting with baryonic matter through Higgs boson exchange is introduced to investigate the combined effects of these two microscopic degrees of freedom on the macroscopic properties of neutron stars. The allowed dark matter parameter space for each model is then progressively constrained using mass-radius observations of neutron stars across different mass ranges and the tidal deformability constraints from GW170817. The nonradial $f$-mode oscillation frequencies are further calculated for the models that remain consistent with these observational constraints.

We first examine the effects of uncertainties in hyperon interactions on neutron star structure. Different hyperon-meson coupling schemes significantly modify the EOS at high densities and consequently affect the maximum neutron star mass, whereas the differences in the mass-radius relations among the models are relatively small in the low-mass region relevant to HESS J1731$-$347. Therefore, varying hyperon interactions alone is insufficient for all six models to simultaneously satisfy the observational constraints on low-mass and massive neutron stars. Upon introducing fermionic dark matter, the stellar mass-radius relations change appreciably, allowing all six EOSs to satisfy the mass-radius constraints from HESS J1731$-$347 within certain ranges of the dark matter Fermi momentum. However, these observations alone are insufficient to effectively distinguish between different hyperon-meson coupling schemes. Further incorporating the NICER observations of PSR J0030+0451 and PSR J0740+6620 excludes EOS3 and EOS4, while EOS1, EOS2, EOS5, and EOS6 retain parameter ranges consistent with the observational constraints. The maximum mass constraint subsequently narrows the allowed ranges of the four remaining models, whereas the tidal deformability constraint from GW170817 imposes no additional restrictions. The resulting allowed ranges of the dark matter Fermi momentum are $0.029\leq k_F^\chi\leq0.030~\mathrm{GeV}$, $0.028\leq k_F^\chi\leq0.032~\mathrm{GeV}$, $0.027\leq k_F^\chi\leq0.040~\mathrm{GeV}$, and $0.027\leq k_F^\chi\leq0.031~\mathrm{GeV}$ for EOS1, EOS2, EOS5, and EOS6, respectively.

We further investigate the quadrupolar $f$-mode oscillation frequencies of the four remaining models. The results show that, at a fixed dark matter Fermi momentum, the influence of different hyperon-meson coupling schemes on the oscillation frequencies exhibits a pronounced dependence on stellar mass. The oscillation frequencies of the four models are relatively close for canonical neutron stars, whereas the differences gradually increase with increasing stellar mass. At $2.0M_\odot$, the maximum difference in the $f$-mode oscillation frequencies among the four models is approximately $173~\mathrm{Hz}$. On the other hand, increasing the dark matter Fermi momentum leads to higher $f$-mode frequencies in all four models. For EOS5, within the observationally allowed parameter range, the oscillation frequencies increase by approximately $250~\mathrm{Hz}$ and $413~\mathrm{Hz}$ for neutron stars with masses of $1.4M_\odot$ and $2.0M_\odot$, respectively. This indicates that the effect of the dark matter Fermi momentum on the oscillation frequencies also depends on stellar mass and becomes more pronounced in massive neutron stars. Even when all four models satisfy current mass-radius and tidal deformability constraints, their $f$-mode oscillation frequencies can still exhibit noticeable differences, potentially providing complementary information for probing the internal composition of neutron stars and constraining dark matter parameters.

These results demonstrate that, although considerable uncertainties remain in hyperon interactions in high-density matter and the possible presence of dark matter in neutron star interiors, current multimessenger observations can already effectively constrain the allowed parameter space of dark-matter-admixed hyperon star models. Theoretical predictions of $f$-mode oscillation frequencies provide further dynamical information distinct from that obtained from mass-radius relations and tidal deformability. With the increasing sensitivity of third-generation gravitational-wave detectors, neutron star oscillation signals can provide new observational information to distinguish between different internal compositions and further elucidate the effects of hyperon interactions and possible dark matter components on neutron star structure.
\section{Acknowledgments}
This work is supported by the Sichuan Science and Technology Program (Grants Nos. 2026YFX0027 and 27ZRMS2773), and the Fundamental Research Funds of China West Normal University (Grant No. 24kx005).
\bibliographystyle{spphys}
\bibliography{template}


\end{document}